\pdfoutput=1 
 \documentclass[preprint,3p,times]{elsarticle}
\usepackage{amssymb}
 \usepackage{amsthm}
 \usepackage{amsmath}
 \usepackage{epstopdf}
\usepackage{siunitx}
\usepackage{graphicx}
\usepackage{pbox}
\usepackage{makecell}

\usepackage[normalem]{ulem}

\begin{document}

\begin{frontmatter}

%% Title, authors and addresses

%% use the tnoteref command within \title for footnotes;
%% use the tnotetext command for theassociated footnote;
%% use the fnref command within \author or \address for footnotes;
%% use the fntext command for theassociated footnote;
%% use the corref command within \author for corresponding author footnotes;
%% use the cortext command for theassociated footnote;
%% use the ead command for the email address,
%% and the form \ead[url] for the home page:
%% \title{Title\tnoteref{label1}}
%% \tnotetext[label1]{}
%% \author{Name\corref{cor1}\fnref{label2}}
%% \ead{email address}
%% \ead[url]{home page}
%% \fntext[label2]{}
%% \cortext[cor1]{}
%% \address{Address\fnref{label3}}
%% \fntext[label3]{}

\title{Contact resistance extraction of graphene FET technologies based on individual device characterization}

%% use optional labels to link authors explicitly to addresses:
%% \author[label1,label2]{}
%% \address[label1]{}
%% \address[label2]{}

\author{Anibal Pacheco-Sanchez, Pedro C. Feijoo, David Jiménez}
\ead{AnibalUriel.Pacheco@uab.cat, PedroCarlos.Feijoo@uab.cat, david.jimenez@uab.cat}
\address{Departament d'Enginyeria Electr\`{o}nica, Escola d'Enginyeria, Universitat Aut\`{o}noma de Barcelona, Bellaterra 08193, Spain.}

\begin{abstract}
%% Text of abstract
Straightforward contact resistance extraction methods based on electrical device characteristics are described and applied here to graphene field-effect transistors from different technologies. The methods are an educated adaptation of extraction procedures originally developed for conventional transistors by exploiting the drift-diffusion-like transport in graphene devices under certain bias conditions. In contrast to other available approaches for contact resistance extraction of graphene transistors, the practical methods used here do not require either the fabrication of dedicated test structures or internal device phenomena characterization. The methodologies are evaluated with simulation-based data and applied to fabricated devices. The extracted values are close to the ones obtained with other more intricate methodologies. Bias-dependent contact and channel resistances studies, bias-dependent high-frequency performance studies and contact engineering studies are enhanced and evaluated by the extracted contact resistance values. 
\end{abstract}
%%%Graphical abstract
%\begin{graphicalabstract}
%%\includegraphics{grabs}
%\end{graphicalabstract}

%%Research highlights
%\begin{highlights}
%\item Contact resistances of graphene transistors are extracted with adapted $Y$-function methods
%\item Bias dependent and independent contact resistance values can be obtained
%\item No additional test structures are required for the extraction
%\item Results are in agreement with physics-based and other experimental approaches
%\item High-frequency performance highly degraded by non-optimal contacts
%\end{highlights}

\begin{keyword}
%% keywords here, in the form: keyword \sep keyword
graphene transistor \sep contact resistance \sep extraction method
%% PACS codes here, in the form: \PACS code \sep code

%% MSC codes here, in the form: \MSC code \sep code
%% or \MSC[2008] code \sep code (2000 is the default)

\end{keyword}

\end{frontmatter}

%% \linenumbers

%% main text
\section{Introduction}
\label{ch:intro}

Graphene (G) field-effect transistors (FETs) have been demonstrated to be suitable candidates for low-power high-frequency (HF) applications in both rigid and flexible substrates \cite{Sch13}. Despite the early stage of this technology, extrinsic cutoff and maximum oscillation frequencies of tens of \SI{}{\giga\hertz} have already been reported in fabricated GFETs  \cite{GuoDon13}-\cite{PenWan19}. In addition to other technological issues to be overcome, the metal-graphene interface in GFETs needs to be further optimized towards exploiting the graphene intrinsic properties, e.g., high velocity saturation and high mobility, at a device level, towards improving static and dynamic device characteristics \cite{GiuDiB17}-\cite{XiaPer11}

In general, the contact resistance in GFETs is a representation of the physical mechanisms preventing the current flow at the interface between the metal contacts and the graphene channel. A correct and efficient characterization of this parameter is a critical point for the development of this emerging technology. A sophisticated physical description of the metal-graphene interface is preferred for the understanding of the carrier injection processes \cite{XiaPer11,ChaJim15}, however, this might be unsuitable for an immediate device characterization since internal device quantities are required in this approach. As an alternative, analytical and compact device models are able to describe specific GFETs by using certain fitting parameters, including the contact resistance  \cite{LanJim14}-\cite{MavWei19}. However, these are  technology-specific approaches which rely on the physical basis of the models and on the calibration procedure.

Test-structure-based characterization methods are of more practical use in laboratories than the modeling approaches for contact resistance assessment. Some of them such as the two-point/four-point-measurement (2P/4P) technique \cite{GuoDon13}, \cite{NagTor11}, the cross bridge Kelvin method \cite{NagTor11}, \cite{NagNis10} and the widely used transfer length method (TLM) \cite{LyuLu16,XiaPer11,SmiVaz15,AnzMan18,VenDri18} have been used in GFETs. However, they involve the fabrication of additional devices and/or special measurement setups and hence, they represent a higher-cost solution in fabrication terms and a less straightforward option for immediate characterization purposes. Furthermore, the reliability of some of these methodologies for graphene FETs is still an open discussion at this early stage of the technology \cite{GiuDiB17,VenDri18,SmiWag17_I}.

Practical and efficient extraction methods in which contact resistance values can be obtained from individual transistor characteristics are required in order to ease device and technology evaluation. In this work, two $I-V$-based methodologies, enabled by a drift-difussion description of the transport in GFETs, are presented and applied to simulation and experimental data.

\section{GFET contact resistances} \label{ch:con}

The device total resistance $R_{\rm{tot}}(=V_{\rm{DS}}/I_{\rm{D}})$ of a GFET embraces the channel resistance $R_{\rm{ch}}$ due to scattering processes and/or defects in the graphene layer(s) and the resistances associated to the source and drain contact regions, $R_{\rm{C,S}}$ and $R_{\rm{C,D}}$, respectively. The latter can be lumped into a total contact resistance $R_{\rm{C}}(=R_{\rm{C,S}}+R_{\rm{C,D}})$, i.e., $R_{\rm{tot}}=R_{\rm{ch}}+R_{\rm{C}}$.

In general, two transport processes occur in a metal-graphene interface: from the metal contact to the coated-graphene region and from the coated-graphene region to the uncoated-graphene region \cite{XiaPer11}, \cite{ChaJim15}, \cite{KonRuh18}. The contact material \cite{ChaJim15}, \cite{AnzMan18}, the contact geometry and dimensions \cite{NagTor11}, \cite{AnzMan18}, \cite{KonRuh18} as well as possible additional layers between metal and graphene \cite{HsuWan11} have an impact on the resistance originated by the first process. A bias-dependent potential barrier induced by a difference in the electronic properties of the coated and the uncoated graphene portions \cite{XiaPer11}, \cite{ChaJim15} is the main cause of the resistance associated to the second process. By considering a practical point of view, these resistances are embraced in this work by the contact resistance corresponding to the drain or source contact in a GFET. Notice that, from a modeling point of view, the impact of the potential barrier at the metal-graphene interface on the performance of Schottky-type devices, such as GFETs, can be considered either into the channel resistance or into a bias-dependent
contact resistance. $R_{\rm{C}}$ extracted at a single bias point in the device linear operation regime is generally provided for technology evaluation \cite{HsuWan11}, \cite{MavWei19}, \cite{AnzMan18}, however, a bias-dependent $R_{\rm{C}}$ reveals more information on internal physical phenomena at the metal-graphene interface.

\section{Y-function-based contact resistance extraction methods for GFETs} \label{ch:extract}

Graphene transistors of different channel and gate lengths have been successfully described by a drift-diffusion (DD) approach \cite{MerHan08}-\cite{WanJin16} due to unavoidable scattering centers deviating the carrier transport within the channel from ideal ballistic conditions. Furthermore, mobility models inspired by conventional Si theory have described GFET experimental results \cite{DorBae10}, \cite{WanJin16}. Extraction methods for contact resistance of GFETs, based on drift-diffusion theory, are presented next.

The $Y$-function \cite{Ghi88} describes a relation of a DD drain current $I_{\rm{D}}$ equation at the linear region and its corresponding transconductance $g_{\rm{m}}(=\partial I_{\rm{D}}/\partial V_{\rm{GS}})$ such as $Y=I_{\rm{D}}/\sqrt{g_{\rm{m}}}$, where the impact of mobility reduction effects has been removed \cite{Ghi88}. Straightforward $Y$-function-based methodologies (YFMs) have been adapted \cite{PacCla16}, \cite{PacCla20} and applied \cite{PacJim19} for device parameters extraction, including $R_{\rm{C}}$, of emerging transistor technologies. In order to consider YFM for GFETs, the underlying transport equation needs to embrace the physical phenomena associated to graphene devices, e.g., Dirac-cone bandstructure  \cite{MerHan08}-\cite{Anc10}. 

%Notice that $R_{\rm{C}}$ values extracted with YFMs have been proven to be more accurate if minimum simplifications of the DD current model are considered \cite{PacCla16}.

%which have been proven useful also in other emerging transistor technologies \cite{PacCla16}, \cite{PacCla20}.

By assuming that the electron carrier transport in GFETs can be described by the DD-approach at the linear unipolar (ohmic) operation regime, and assuming that the carrier concentration can be computed as the average between the charge at the source side $C_{\rm{ox}}(V_{\rm{GS}}-V_{\rm{Dirac}})$ and the charge at the drain side $C_{\rm{ox}}(V_{\rm{GD}}-V_{\rm{Dirac}})$, $I_{\rm{D}}$ is given by \cite{PacCla16}, \cite{PacCla20}

\begin{equation} 
I_{\rm{D}} \approx \beta \frac{\left(V_{\rm{GS}} - V_{\rm{Dirac}} - \frac{V_{\rm{DS}}}{2} \right)}{1+\theta \left(V_{\rm{GS}} - V_{\rm{Dirac}} - \frac{V_{\rm{DS}}}{2} \right) } V_{\rm{DS}},
\label{eq:YFM_Id}
\end{equation}

\noindent where $V_{\rm{GS}/\rm{DS}}$ is the extrinsic gate-to-source/drain-to-source voltage, $V_{\rm{Dirac}}=V_{\rm{GS}}\vert_{\rm{min}(\mathit{I}_{\rm{Dirac}})}\sim V_{\rm{GS0}}+V_{\rm{DS}}/2$ is the Dirac voltage with $V_{\rm{GS0}}$ as the flat-band voltage \cite{FeiPas19}, $\beta=\mu_0 C_{\rm{ox}}w_{\rm{g}}/L_{\rm{g}}$ with a low-field mobility $\mu_0$, the oxide capacitance $C_{\rm{ox}}$, the gate width $w_{\rm{g}}$ and length $L_{\rm{g}}$, and $\theta$ is the \textit{extrinsic} mobility attenuation factor $\theta=\theta_0 + R_{\rm{C}}\beta$ \cite{Ghi88}, \cite{HaoCab85} with the \textit{instrinsic} attenuation factor due to vertical fields $\theta_0$. 
%Notice that Eq. (\ref{eq:YFM_Id}) is a rough approximation of the transport in GFETs useful for device parameter extraction and only valid in a strong inversion regime-like of the device. For modeling purposes in all-transistor operation regions, other more thorough approaches should be followed \cite{MerHan08}-\cite{WanJin16}.

By considering Eq. (\ref{eq:YFM_Id}), the corresponding $Y$-function yields

\begin{equation}
Y = \sqrt{\beta V_{\rm{DS}}} \left(V_{\rm{GS}}-V_{\rm{Dirac}} - \frac{V_{\rm{DS}}}{2} \right),
\label{eq:Yfun_original}
\end{equation}

\noindent from where $\beta$ can be obtained at the maximum point of its derivative $Y'$ with respect to $V_{\rm{GS}}$ for each $V_{\rm{DS}}$. The maximum derivative has been choosen in order to guarantee a linear operation limit. Similarly, a function $X=1/\sqrt{g_{\rm{m}}}$ is given by

\begin{equation}
X = \frac{1 + \theta \left(V_{\rm{GS}} - V_{\rm{Dirac}} - \frac{V_{\rm{DS}}}{2} \right)}{\sqrt{\beta V_{\rm{DS}}}},
\label{eq:X_fun}
\end{equation}

\noindent the derivative of which yields a value for $\theta$ at the maximum point of its derivative $X'$ with respect to $V_{\rm{GS}}$ for each $V_{\rm{DS}}$. A $V_{\rm{GS}}$-independent contact resistance value $R_{\rm{C,1}}$ is extracted from the slope of the relation of $\theta$ with respect to $\beta$, once these terms have been obtained as described above, i.e., $R_{\rm{C,1}} = \partial \theta/\partial \beta$.

Alternatively, in order to obtain a $V_{\rm{GS}}$-dependent contact resistance $R_{\rm{C,2}}$, an expression can be obtained by applying the definition of $\theta$ in Eq. (\ref{eq:YFM_Id}) and using Eqs. (\ref{eq:Yfun_original}) and (\ref{eq:X_fun}) for rearranging terms. $R_{\rm{C,2}}$ is hence given by \cite{PacCla20}

  \begin{equation}
  \begin{aligned}
R_{\rm{C,2}} = \frac{V_{\rm{DS}}}{Y^2}  & \left( V_{\rm{GS}} - V_{\rm{Dirac}} - \frac{V_{\rm{DS}}}{2} \right)^2 \cdot \\ & \left[  \frac{XY - 1}{\left(V_{\rm{GS}}-V_{\rm{Dirac}}-\frac{V_{\rm{DS}}}{2}\right)^2} - \theta_0 \right].
%\left[ \left( \frac{XY}{V_{\rm{GS}}-V_{\rm{Dirac}}-\frac{V_{\rm{DS}}}{2}} - 1 \right) \left( \frac{1}{V_{\rm{GS}}-V_{\rm{Dirac}} - \frac{V_{\rm{DS}}}{2}} \right) - \theta_0 \right].
    \label{eq:Rc_YFM_Vg}
    \end{aligned}
  \end{equation}

According to the methods' features, the transfer characteristics at different $V_{\rm{DS}}$, rather than the output characteristics of the device are required for the $R_{\rm{C}}$ extraction. The methods can be also applied to hole transport by properly adapting Eq. (\ref{eq:YFM_Id}) to a hole drain current model and following a similar approach as described above. In contrast to the widely used TLM, no assumption of uniform sheet resistance along the channel and the region under the contacts \cite{Sch06} is required in YFM. In order to obtain reliable reproducible values, the impact of unavoidable traps is recommended to be reduced in experimental data, e.g., by pulsed measurements \cite{HafPac16}, \cite{CarSer14}. The extracted contact resistance values are useful for practical applications since they correspond to the bias region where GFETs are expected to work in HF circuits. In contrast to a previous study where $R_{\rm{C}}$ of GFETs has been extracted with a different YFM \cite{UrbLup20}, the DD curent model here involves minimum simplifications which has been proven to yield more accurate results in carbon-based devices \cite{PacCla16}. Furthermore, a practical difference with the method in \cite{UrbLup20} is that for the extraction of $R_{\rm{C}}$ here, the characterization of $C_{\rm{ox}}$ is not required.

\section{Contact characterization of different GFET technologies} \label{ch:exp}

\subsection{Simulated devices}

\begin{figure}[!htb]
\centering
\includegraphics[height=0.245\textwidth]{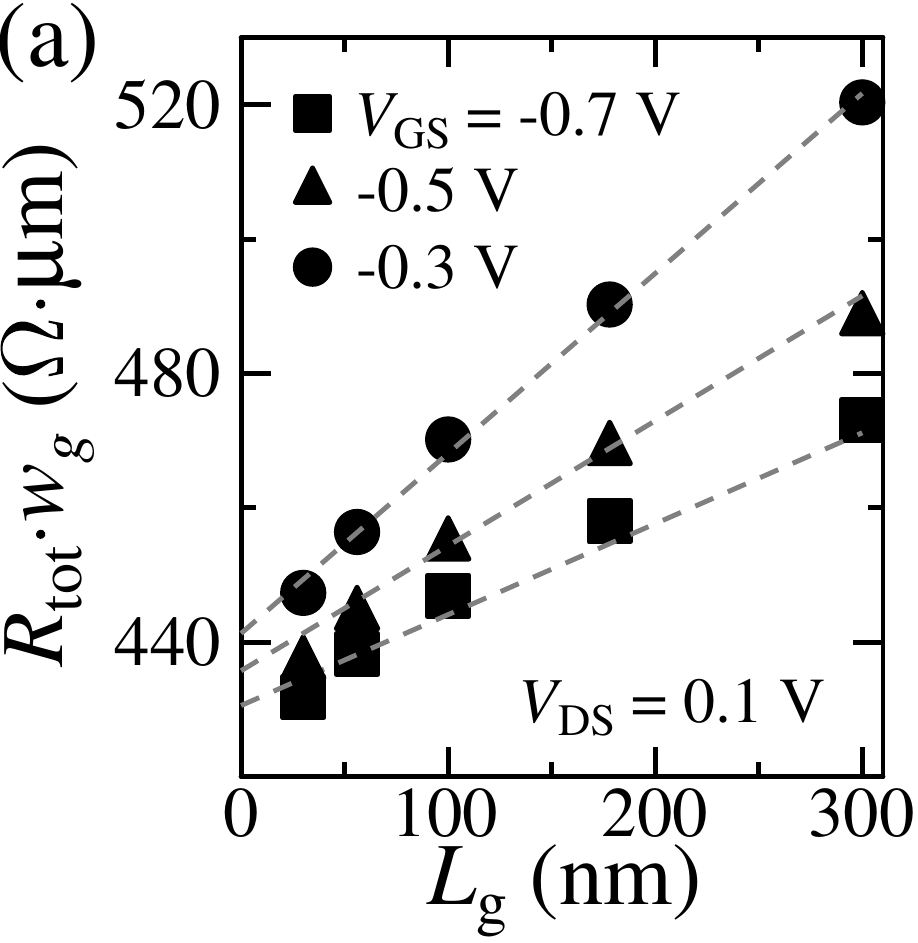}
\includegraphics[height=0.245\textwidth]{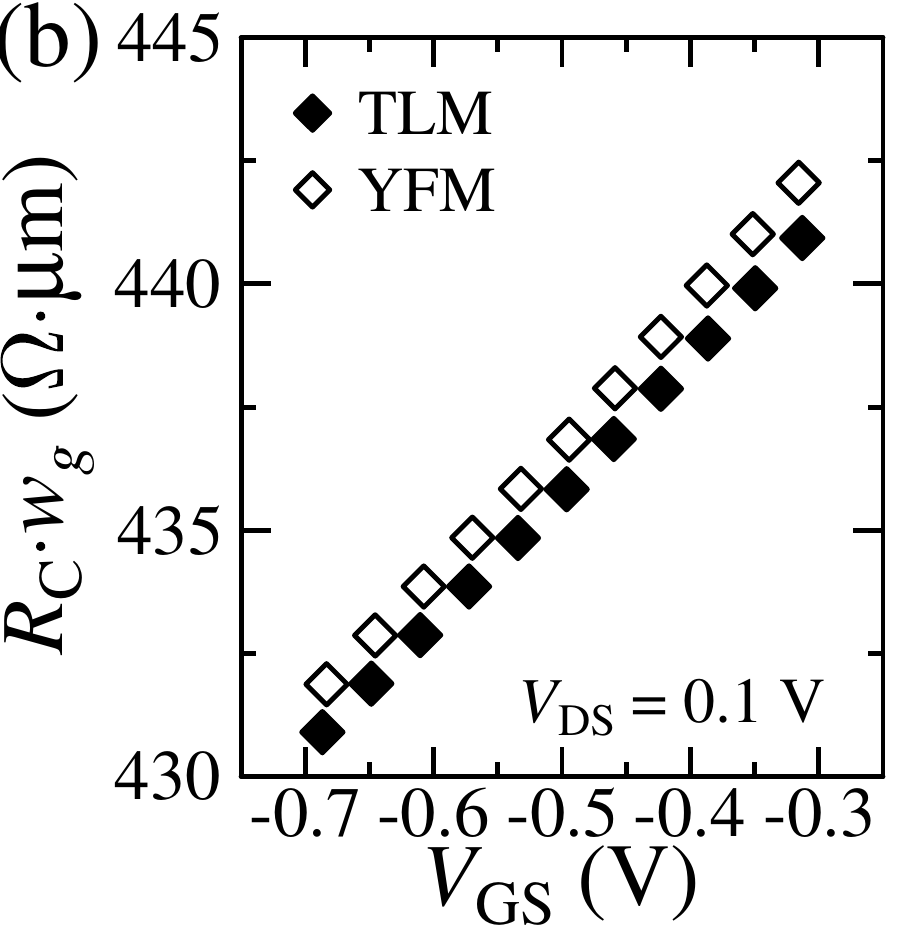} \\
\caption{(a) TLM plot obtained from simulated devices at different bias. Dashed lines correspond to a linear fitting of each curve. (b) Contact resistivity of simulated devices extracted with TLM and YFM presented here (Eq. (\ref{eq:Rc_YFM_Vg})).}
\label{fig:TLM}
\end{figure}

Scattering-affected DC transfer characteristics of top-gate GFETs with identical device architecture but with different gate lengths have been generated with numerical device simulations consisting on a self-consistent solution of the Poisson's equation and the current-continuity equation \cite{FeiPas19}. This set of different $L_{\rm{g}}$ devices enables to imitate a TLM structure.  The simulated devices consist of an hexagonal boron nitride (h-BN) encapsulated graphene channel with top and bottom h-BN layers thicknesses of \SI{30}{\nano\meter} each and relative permittivity of \SI{3}{} and a \SI{285}{\nano\meter} SiO$_2$ substrate layer. The mobility is assumed to decrease with the vertical field. Further details on the considerations made for device geometry and carrier mobility can be found in \cite{FeiPas19}. A reference constant contact resistivity of \SI{400}{\ohm\cdot\micro\meter} associated to interface layers has been set. The devices gate lengths are of \SIlist{30;56;100;178;300}{\nano\meter}. The corresponding TLM curves obtained at different bias are shown in Fig. \ref{fig:TLM}(a).

The contact resistivity $R_{\rm{C}}\cdot w_{\rm{g}}$, extracted with Eq. (\ref{eq:Rc_YFM_Vg}), for the \SI{100}{\nano\meter}-long simulated device\footnote{Results of other devices under study have similar trend and magnitude of values (not shown here).} is in good agreement with results of the same parameter obtained via TLM within the same bias region as shown in Fig. \ref{fig:TLM}(b). The $V_{\rm{GS}}$-dependence and the larger value of extracted $R_{\rm{C}}\cdot w_{\rm{g}}(>\SI{400}{\ohm\cdot\micro\meter})$ in comparison to the reference value indicate that both extraction methods embrace not only the impact of interfacial layers but also phenomena associated to the internal bias-dependence potential barrier (see Section \ref{ch:con}).

The extraction methods have also been applied to data from a physics-based model (Landauer transport theory) \cite{ChaJim16} of a graphene-based transistor-like device (see Fig. 10 in \cite{ChaJim16}) with a reference contact resistance in the model $R_{\rm{C,Land}} = R_{\rm{C,PB,ref}} + R_{\rm{C,IL}}$. $R_{\rm{C,Land}}$ in  \cite{ChaJim16} embraces a resistance associated to a potential barrier $R_{\rm{C,PB,ref}}$ of \SI{250}{\kilo\ohm} and a resistance associated to other interfacial layers $R_{\rm{C,IL}}$ (value not reported). Notice that a metal/oxide/graphene/semiconductor stack is considered within the contact region and that the oxide is isotropic and isometric along the whole device in the direction of the carrier transport including the gated-graphene region under the top-gate \cite{ChaJim16}. $R_{\rm{C,1}}$ has been extracted for devices in \cite{ChaJim16} with identical architectures but different oxide thickness $t_{\rm{ox,MG}}$ at the metal-graphene interface: a \SI{10}{\nano\meter} $t_{\rm{ox,MG}}$ for the simulated device under study (SDUT) A and a \SI{100}{\nano\meter} $t_{\rm{ox,MG}}$ for SDUT-B.

%\begin{table} [!htb] 
%\begin{center}
%\caption{Values of reported ($R_{\rm{C,SB,ref}}$) and extracted ($R_{\rm{C,1}}$, $R_{\rm{ch,ext}}$ and $R_{\rm{C,IL,ext}}$) resistances of simulated graphene-based barristors}
%\begin{tabular}{c|c||c|c|c}
%
%\makecell{$t_{\rm{ox}}$\\ (\si{\nano\meter})} & \makecell{$R_{\rm{C,PB,ref}}$\\ (\si{\kilo\ohm})} & \makecell{$R_{\rm{C,1}}$\\ (\si{\kilo\ohm})} & \makecell{$R_{\rm{ch,ext}}$\\ (\si{\kilo\ohm})} & \makecell{${R_{\rm{C,IL,ext}}}$\\ (\si{\kilo\ohm})} \\ \hline \hline
%
%\SI{10}{}  & \SI{250}{} & \SI{342}{} & \SI{76}{}  & \SI{92}{} \\
%\SI{100}{} & \SI{250}{} & \SI{351}{} & \SI{397}{} & \SI{101}{} \\
%
%\end{tabular} \label{tab:rc_Land}
%\end{center}
%\end{table}

A smaller oxide improves the gate control over the graphene channel and hence the potential barrier is reduced \cite{ChaJim16}. The latter is the same tendency observed from the extracted smaller $R_{\rm{C,1}}$ of \SI{342}{\kilo\ohm} for SDUT-A in contrast to the \SI{351}{\kilo\ohm} extracted for SDUT-B. From $R_{\rm{C,1}}(=R_{\rm{C,PB,ref}} + R_{\rm{C,IL,ext}})$ on this study, the  increase of the interface-layers resistance $R_{\rm{C,IL,ext}}$ due to a larger $t_{\rm{ox,MG}}$ can be also observed since a \SI{101}{\kilo\ohm} for SDUT-B has been extracted in contrast to the \SI{92}{\kilo\ohm} obtained for SDUT-A. Furthermore, the larger increase of the extracted channel resistance $R_{\rm{ch,ext}}(=R_{\rm{tot}}-R_{\rm{C,1}})$, associated to a higher number of scattering events in the thicker device (\SI{76}{\kilo\ohm} for SDUT-A, \SI{397}{\kilo\ohm} for SDUT-B), in contrast to the almost similar $R_{\rm{C,1}}$ in both cases, indicate that the channel phenomena have no impact on the extraction method. 

%A thicker oxide leads to higher $R_{\rm{C,iL,ext}}$ due to the larger path the carriers need to follow in order to participate in the current transport. 

\subsection{Fabricated GFET technologies}

\begin{figure}[!htb]
\centering
\includegraphics[height=0.245\textwidth]{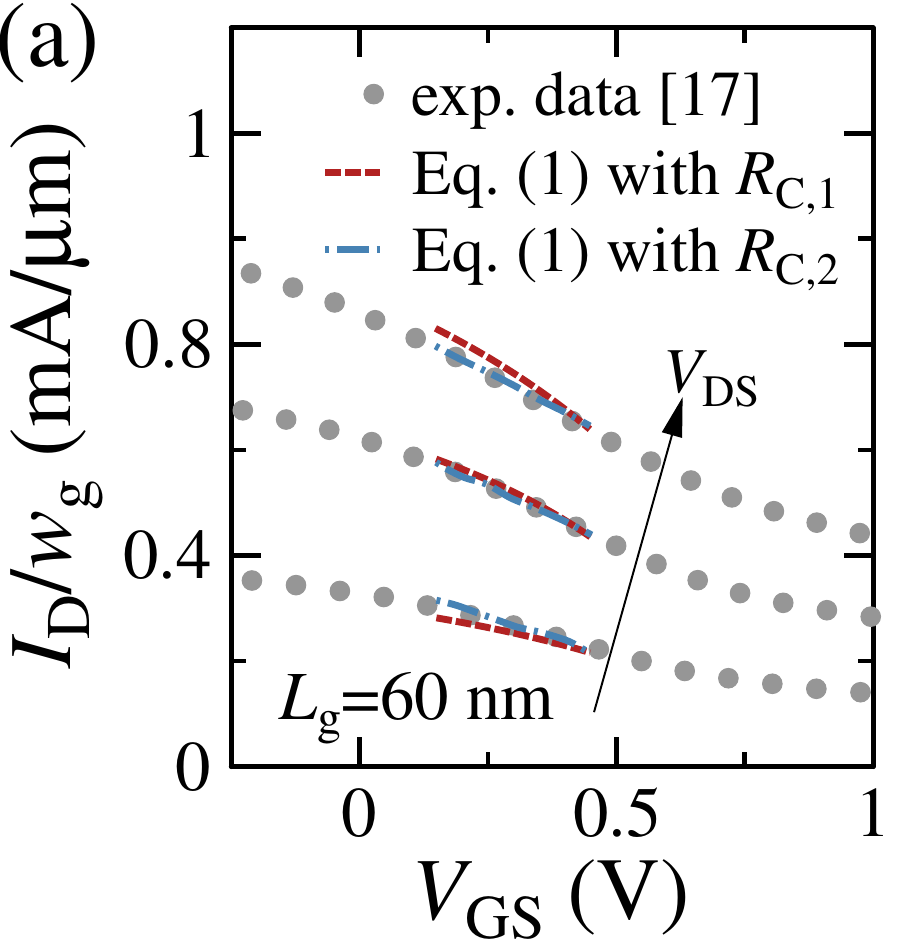}
\includegraphics[height=0.245\textwidth]{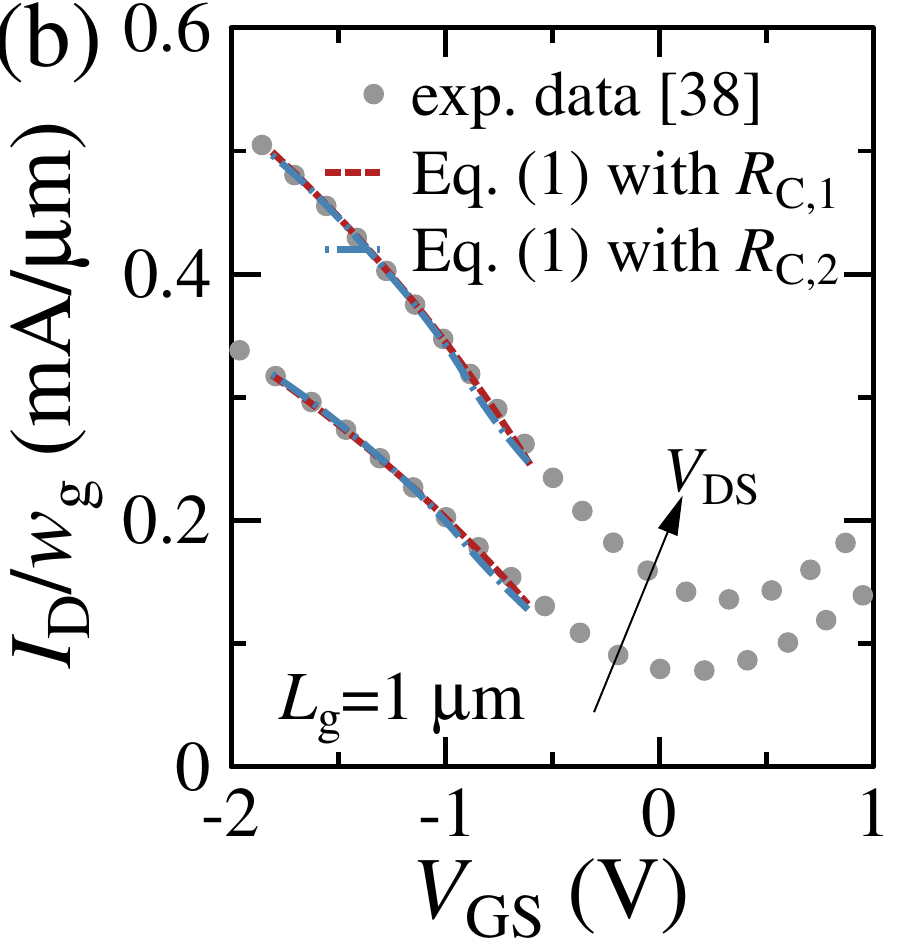} \\
\caption{Transfer characteristics of different fabricated GFET technologies: (a) $w_{\rm{g}}/L_{\rm{g}}=\SI{20/60}{\micro\meter/\nano\meter}$ \cite{WuZou16} at $V_{\rm{DS}}$ equal to \SIlist{0.1;0.2;0.3}{\volt} and (b) $w_{\rm{g}}/L_{\rm{g}}=\SI{20/1000}{\micro\meter/\nano\meter}$ \cite{TiaLi18} at $V_{\rm{DS}}$ equal to \SIlist{1;1.5}{\volt}. Symbols represent experimental data and lines correspond to results with Eq. (\ref{eq:YFM_Id}) considering the extracted parameters.}
\label{fig:IdVg_exa}
\end{figure}

Contact resistance values have been extracted with the methods discussed above for a wide variety of GFET technologies \cite{GuoDon13,LyuLu16}, \cite{HsuWan11}, \cite{SmiVaz15}-\cite{MavWei19}, \cite{TiaLi18}-\cite{KedHsu09}, i.e., devices with different footprints, architectures and fabrication processes have been considered. The extractions have been performed here considering the dominant branch of the transfer characteristic in each device. Eq. (\ref{eq:YFM_Id}) has been calculated with the extracted parameters, including $R_{\rm{C,1}}$ or $R_{\rm{C,2}}$ -depending on the selected method-, for each device under study and the results have been compared to the corresponding experimental data. A good match between such curves validates the parameters within the bias range selected for the extraction. This validation procedure has been applied for all devices. As an example, Fig. \ref{fig:IdVg_exa} shows the good match between experimental data of devices with different gate lengths (\SI{60}{\nano\meter} \cite{WuZou16} and \SI{1}{\micro\meter} \cite{TiaLi18}) and Eq. (\ref{eq:YFM_Id}) using the corresponding extracted parameters, including the contact resistance. Additional curves for different GFET technologies (\cite{LyuLu16}, \cite{WanHsu11}), i.e., different geometries, have been presented elsewhere \cite{PacJim19} with similar results. Notice that the $I-V$-based verification procedure presented here has not been performed for the contact resistance values extracted with other methods in the corresponding reference.

\begin{figure}[!htb]
\centering
\includegraphics[height=0.245\textwidth]{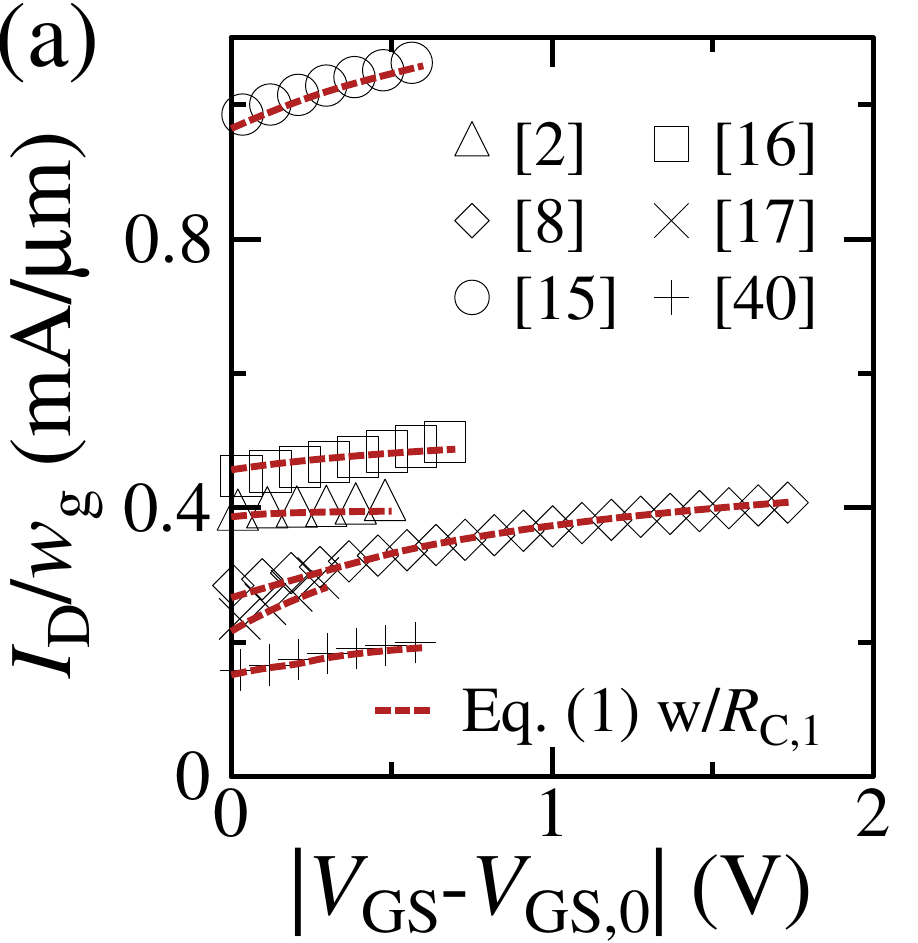}
\includegraphics[height=0.245\textwidth]{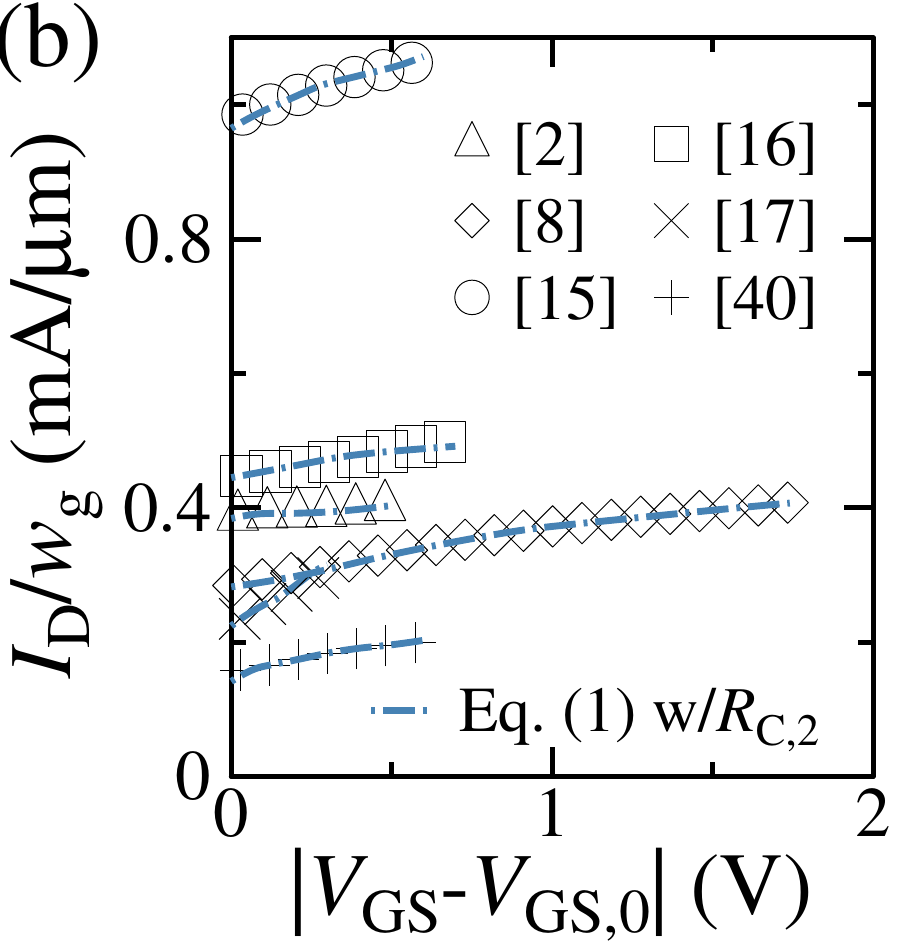} \\
\caption{Transfer characteristics within the bias range where the extraction methods have been applied of devices with different gate lengths: \SI{60}{\nano\meter} \cite{WuZou16}, \SI{150}{\nano\meter} \cite{SanXu18}, \SI{150}{\nano\meter} \cite{MelFre13}, \SI{250}{\nano\meter} \cite{GuoDon13}, \SI{300}{\nano\meter} \cite{CheBai12} and \SI{2000}{\nano\meter} \cite{HsuWan11}. Symbols represent experimental data and lines represent Eq. (\ref{eq:YFM_Id}) using the extracted (a) $R_{\rm{C,1}}$ and (b) $R_{\rm{C,2}}$.}
\label{fig:IdVg_all}
\end{figure}

In order to further demonstrate the validity of the extraction methods presented above, the experimental transfer characteristics of the different GFET technologies and their corresponding description with Eq. (\ref{eq:YFM_Id}) using the extracted parameters, including $R_{\rm{C,1}}$ and $R_{\rm{C,2}}$ are shown in Fig. \ref{fig:IdVg_all} where $\vert V_{\rm{GS,0}} \vert$ is the lowest gate-to-source voltage in which the methods have been applied. Results are shown at the lowest reported $\vert V_{\rm{DS}} \vert$ in all studies in order to ease the discussion, however, similar accurate descriptions have been also obtained at different $\vert V_{\rm{DS}} \vert$ in all cases (not shown here).
 
Eq. (\ref{eq:YFM_Id}) correctly describes the experimental data of all technologies under study using the extracted values, including $R_{\rm{C,1}}$ and $R_{\rm{C,2}}$, from both methodologies. The extraction methods have been applied  here within a bias region ($\vert V_{\rm{GS}}- V_{\rm{GS,0}} \vert$) of interest for HF circuit applications, such as the strong linear regime, i.e., the bias corresponding to the Dirac point has not been considered. Due to noisy data and diverse technology-related effects, the bias range in which Eq. (\ref{eq:YFM_Id}) is valid differs from device to device and thus, the extraction methods have been applied accordingly.

The contact resistivity values, extracted using $R_{\rm{C,1}}$, of the different evaluated technologies \cite{GuoDon13,LyuLu16}, \cite{HsuWan11}, \cite{SmiVaz15}-\cite{MavWei19}, \cite{TiaLi18}-\cite{KedHsu09} are listed in Table \ref{tab:rc_exp}. $R_{\rm{C}}\cdot w_{\rm{g}}$ values of the same devices reported with test-structured based methods (TLM and 2P) and analytical model (AM) or compact model (CM) calibration are included as well in Table \ref{tab:rc_exp} for comparison along with some device geometry parameters of the studied technologies. $R_{\rm{C,1}}$ has been extracted at a similar bias point at which the reference contact resistance $R_{\rm{C,ref}}$ value has been reported in the corresponding study. Despite the universality of the methods described above, a scaling study is not feasible here since the studied devices are from different technologies. 

\begin{table} [!htb] 
\caption{Contact resistivity extracted at a single bias point $R_{\rm{C,1}}\cdot w_{\rm{g}}$, reported values of contact resistivity  $R_{\rm{C,ref}}\cdot w_{\rm{g}}$ obtained with other methods and device dimensions of fabricated GFETs. }
%\begin{indented}
\begin{tabular}{c|c|c||c|c} 

[ref.] & \makecell{$w_{\rm{g}}$\\ (\si{\micro\meter})} & \makecell{${L_{\rm{g}}}$\\ (\si{\nano\meter})} & \makecell{${R_{\rm{C,ref}}}\cdot w_{\rm{g}}$\\ (\si{\kilo\ohm \micro\meter})} & \makecell{${R_{\rm{C,1}}}\cdot w_{\rm{g}}$\\ (\si{\kilo\ohm \micro\meter})} \\ \hline \hline
\multicolumn{5}{c}{with 2P} \\ \hline

\cite{GuoDon13} & \SI{14}{}  & \SI{100}{} & \SI{0.2}{} & \SI{0.2}{} \\ 

\cite{GuoDon13} & \SI{14}{}  & \SI{250}{} & \SI{0.2}{} & \SI{0.2}{} \\ \hline

\multicolumn{5}{c}{with TLM} \\ \hline

\cite{LyuLu16} & --  & \SI{100}{} & \SI{1.1}{} & \SI{1.4}{}   \\ 

\cite{LyuLu16} & --  & \SI{300}{} & \SI{1.1}{} & \SI{1.4}{}   \\  

\cite{SmiVaz15} & \SI{80}{}  & \SI{2000}{} & \SI{20}{} (also w/AM) & \SI{22}{} \\ \hline

\multicolumn{5}{c}{with AM} \\ \hline

\cite{WuZou16} & \SI{20}{}  & \SI{60}{} & \SI{0.2}{} & \SI{0.2}{} \\  %the original Rc should be multiplied by 2 in order to have same definition

\cite{HsuWan11} & \SI{10}{}  & \SI{2000}{} & \SI{1.2}{} & \SI{1.4}{} \\ \hline

\multicolumn{5}{c}{with CM} \\ \hline

\cite{MavWei19} & \SI{12}{}  & \SI{100}{} & \SI{3.1}{} & \SI{3.3}{} \\ 

\cite{SanXu18} & \SI{12}{}  & \SI{150}{} & \SI{0.08}{} & \SI{0.2}{}   \\ 

\cite{MelFre13} & \SI{24}{}  & \SI{150}{} & \SI{2.4}{} & \SI{2.3}{} \\  %at VGS=-2 V , VDS=3V

\cite{MavWei19} & \SI{12}{}  & \SI{300}{} & \SI{6.2}{} & \SI{6.3}{} \\   

\cite{WeiDeo14}  & \SI{12}{}  & \SI{300}{} & \SI{4.6}{} in \cite{AguFre17} & \SI{4.3}{} \\ 

\cite{IanMuz15} & \SI{40}{}  & \SI{4000}{} & \SI{16}{} & \SI{20}{} \\ 

\cite{WanHsu11} & \SI{25}{}  & \SI{5000}{} & \SI{7}{} in \cite{LanJim14} & \SI{10}{} \\

%\cite{HanChe11} & \SI{5}{}  & \SI{5000}{} & \makecell{\SI{1.5}{}\\(with TLM)} & \SI{}{} \\ \hline

\cite{KedHsu09} & \SI{5}{}  & \SI{10000}{} & \SI{3}{} in \cite{JimMol11} & \SI{3.2}{} \\  \hline

\multicolumn{5}{c}{$R_{\rm{C}}$ not extracted previously} \\ \hline

\cite{CheBai12} & \SI{5}{}  & \SI{300}{} & --    & \SI{0.5}{}   \\    

\cite{TiaLi18} & \SI{20}{}  & \SI{1000}{} & -- & \SI{3.4}{} \\ 

\cite{SmiWag17} & \SI{20}{}  & \SI{4000}{} & -- & \SI{52}{} \\ 

%\cite{PasGah18} & -- & -- & -- & $\SI{36e3}{}\cdot w_{\rm{g}}$ \\ \hline

\cite{KonWan17} & \SI{15}{} & \SI{6000}{} & -- & \SI{0.2}{} \\ 
 %check data with Vbg=60V
%\cite{HadHol12} & \SI{20}{}  & \SI{750}{} & \makecell{\SI{0.1}{}\\with ?} & \SI{}{} \\ 
\end{tabular} \label{tab:rc_exp}
\end{table}

%\begin{figure}[!htb]
%\centering
%\includegraphics[height=0.2525\textwidth]{fig/RcWg_Lg_02}
%%\includegraphics[height=0.2025\textwidth]{fig/RchRc_Vg_dev_paper_GFET_pres} \\
%\caption{Comparison of the contact resistivity values of different technologies extracted using $R_{\rm{C,1}}$ and other methods.}
%\label{fig:Rc1Wg}
%\end{figure}
%\cite{GuoDon13,LyuLu16,HsuWan11,SmiVaz15,SanXu18,MelFre13,WuZou16,MavWei19,WeiDeo14,CheBai12,SmiWag17,WanHsu11,IanMuz15,KedHsu09},  
The extracted values are close to the reference ones obtained by other more intricate and technology-specific methods. The mean relative deviation of $R_{\rm{C,1}}$ and reference values, excluding the \SI{150}{\nano\meter}-long device \cite{SanXu18}, is between 6\% and 27\%. However, not all the reference values are validated in contrast to the procedure included here (see Fig. \ref{fig:IdVg_all}). The difference between the extracted and reference data of the \SI{150}{\nano\meter}-long device \cite{SanXu18} can be explained by a strong impact of the Schottky barrier in the device performance which is not considered in the reference parameter in contrast to $R_{\rm{C,1}}$ (and $R_{\rm{C,2}}$) here which embraces the Schottky barrier contribution. 

In contrast to contact resistance values obtained by fitting certain CMs or AMs, the $Y$-function based methods presented here work for different technologies without adjusting further parameters. While physics-based models are more accurate to describe the device performance, they result impractical from the characterization point of view since they require information regarding intrinsic physical device values, e.g., charge carrier density \cite{HsuWan11}, \cite{SmiVaz15}. Therefore, the methods discussed here are an alternative for immediate contact resistance extraction. 

The same value of $R_{\rm{C,1}}$ extracted for scattering-affected transistors with identical architecture and materials but different $L_{\rm{g}}$ \cite{GuoDon13}, \cite{LyuLu16}, where $R_{\rm{ch}}$ is expected to differ, indicates that channel phenomena have no impact on the extraction. The latter can be exploited in devices with more sophisticated channel morphology, such as graphene nanoribbons FETs \cite{PasGah18}, in which channel and contact improvements can be evaluated independently. E.g., the $R_{\rm{C,1}}$ of \SI{36}{\mega\ohm} of such device \cite{PasGah18}, extracted with Eq. (\ref{eq:YFM_Id}), should decrease after a contact engineering but remain the same under only channel pattern treatment. Furthermore, the impact on the contacts quality of an electrostatic doping applied to the \SI{1}{\micro\meter}-long device \cite{TiaLi18} can be observed in the reduction from \SI{3.4}{\kilo\ohm\cdot\micro\meter}, obtained with $R_{C,1}$, of the undoped device, to \SI{1.7}{\kilo\ohm\cdot\micro\meter} of the doped device. The methods presented here are also an effective and immediate tool to evaluate contact engineering techniques in a technology such as the $R_{\rm{C}}$ improvement in fabricated GFETs \cite{KonWan17} due to an optical litography treatment ($R_{\rm{C,1}}=\SI{215}{\ohm}$) in contrast to non-treated contacts ($R_{\rm{C,1}}=\SI{450}{\ohm}$). $R_{\rm{C,2}}$ can indicate such improvement over bias (not shown here) in contrast to the techniques in \cite{KonWan17}.
%
%change in \cite{TiaLi18} after electrostatic doping from 3.4 to 1.7

\begin{figure}[!htb]
\centering
\includegraphics[height=0.245\textwidth]{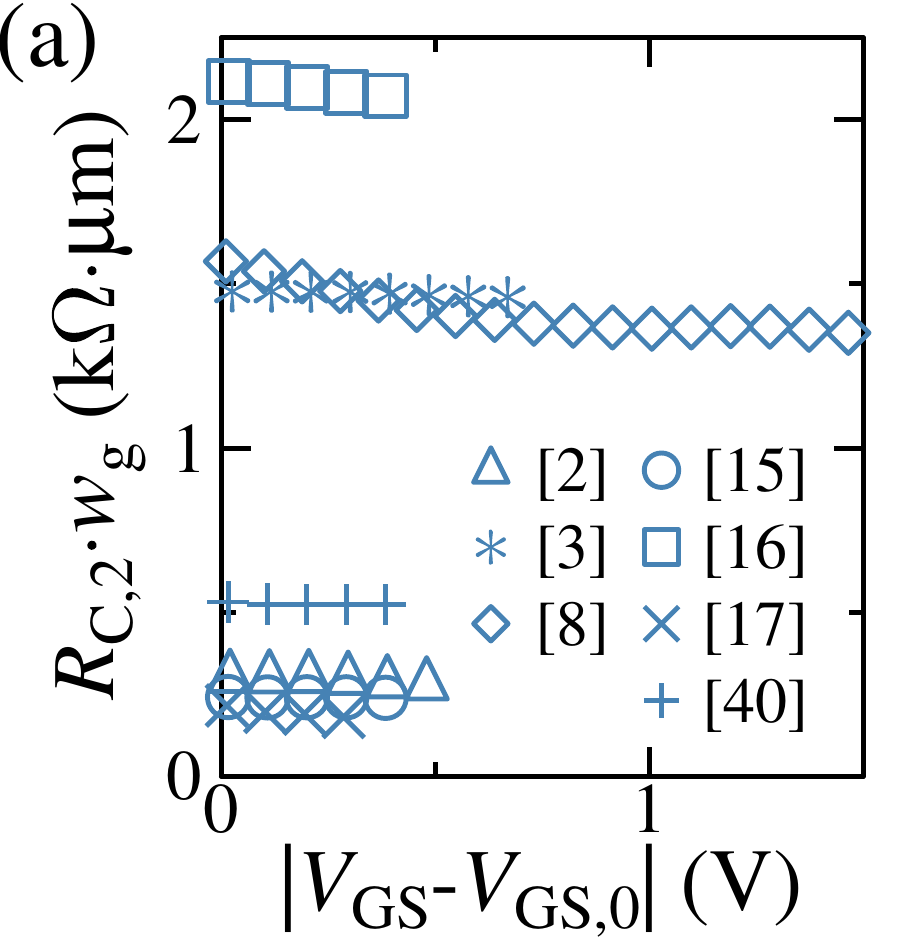}
\includegraphics[height=0.245\textwidth]{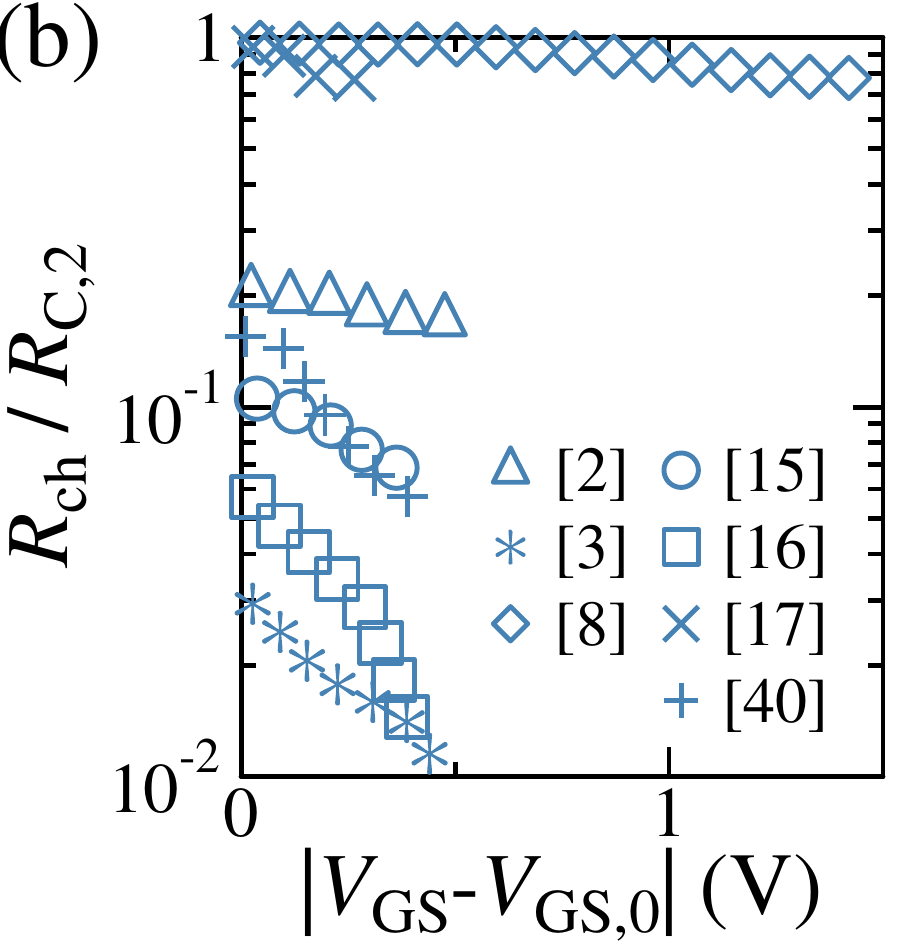} \\
\caption{(a) Extracted $V_{\rm{GS}}$-dependent contact resistivity (using Eq. (\ref{eq:Rc_YFM_Vg})) of fabricated GFETs designed for HF applications \cite{GuoDon13,LyuLu16,HsuWan11,SanXu18,MelFre13,WuZou16,CheBai12}. (b) Ratio between channel resistance and extracted contact resistance $R_{\rm{C,2}}$ for the HF devices. All curves correspond to the lowest reported $V_{\rm{DS}}$.}
\label{fig:Rc2Wg}
\end{figure}

The $V_{\rm{GS}}$-dependent contact resistivity, obtained from Eq. (\ref{eq:Rc_YFM_Vg}), of HF GFETs is presented in Fig. {\ref{fig:Rc2Wg}}(a). Considerations for the bias region in which $R_{\rm{C,2}}$ has been extracted remain the same as above (see discussion of Fig. \ref{fig:IdVg_all}). 

%Results for other devices in Table \ref{tab:rc_exp} have been presented elsewhere \cite{PacJim19}.

A monotonic decrease of $R_{\rm{C,2}}\cdot w_{\rm{g}}$ is observed for an increasing $\vert V_{\rm{GS}}-V_{\rm{GS,0}} \vert$. This is an expected result in the linear operation regime due to the lowering of the potential barrier between metal and graphene. Low values of $R_{\rm{C,2}}\cdot w_{\rm{g}}$ as well as a linear constant response over certain bias, can indicate a suitable contact transparency for the intended low-power HF applications of carbon-based devices \cite{ParAkin11}, \cite{MotCla15}. The device linearity is recommended to be confirmed with trap-free data \cite{HafPac16}, \cite{CarSer14}. The bias-dependent contact resistivity of the other devices under study can be found elsewhere \cite{PacJim19}.

The channel to contact resistance ratio shown in Fig. \ref{fig:Rc2Wg}(b) indicates the impact of these parameters on the device behavior. The high steepness of the curves is due to the $V_{\rm{GS}}$-dependence of $R_{\rm{ch}}$ rather than that of $R_{\rm{C,2}}$. The contact resistance is extremely dominant for most of the devices \cite{GuoDon13,LyuLu16,SanXu18,MelFre13,CheBai12}. The non-linear response of the curves corresponding to the \SI{150}{\nano\meter}-long device \cite{MelFre13} and the \SI{2}{\micro\meter}-long device \cite{HsuWan11} suggest non-trivial internal mechanisms, e.g., transport through higher sub-bands, trap-induced current variations, etc., the discussion of which is out of the scope of this work. A ratio close to unity indicates that both channel and contact resistances are relevant for the \SI{2}{\micro\meter}-long device \cite{HsuWan11} as well as for the \SI{60}{\nano\meter}-long GFET \cite{WuZou16}. The latter reveals a non-intuitive scattering-affected behaviour of short devices.

\subsection{$R_{\rm{C}}$-based high-frequency performance projection}

GFETs HF performance projection studies over bias are enabled by the $R_{\rm{C,2}}$ and by a oftenly used small-signal model approximation \cite{Sch13}-\cite{LyuLu16}, \cite{FeiPas19} where the extrinsic cutoff frequency $f_{\rm{t,e}}$ and the extrinsic maximum oscillation frequency $f_{\rm{max,e}}$ are given by

\begin{equation}
f_{\rm{t,e}} \approx \frac{f_{\rm{t,i}}}{1+g_{\rm{d,i}}R_{\rm{C}}+2\pi f_{\rm{t,i}}C_{\rm{gd,i}}R_{\rm{C}}},
\label{eq:fte}
\end{equation}

\begin{equation}
f_{\rm{max,e}} \approx \frac{f_{\rm{t,e}}}{2\sqrt{g_{\rm{d,i}}\left(R_{\rm{g}}+R_{\rm{C}}\right) + 2\pi f_{\rm{t,e}}R_{\rm{g}}C_{\rm{gd,i}}}},
\label{eq:fmaxe}
\end{equation}

\noindent respectively, where $f_{\rm{t,i}} \approx g_{\rm{m,i}}/\left[2\pi\left(C_{\rm{gs,i}}+C_{\rm{gd,i}}\right)\right]$ is the intrinsic cutoff frequency, $g_{\rm{m,i}}$ the intrinsic transconductance, $g_{\rm{d,i}}$ the intrinsic output conductance, $C_{\rm{gs,i/gd,i}}$ the intrinsic gate-to-source/gate-to-drain capacitance, $R_{\rm{g}}$ the gate resistance and $R_{\rm{C}}$ the contact resistance, corresponding here to $R_{\rm{C,2}}$. For the calculation of $g_{\rm{m,i}}$ and $g_{\rm{d,i}}$, the intrinsic gate-to-source voltage $V_{\rm{GS,i}}\approx V_{\rm{GS}}-I_{\rm{D}}R_{\rm{C}}/2$ and the intrinsic drain-to-source voltage $V_{\rm{DS,i}}\approx V_{\rm{DS}}-I_{\rm{D}}R_{\rm{C}}$ have been considered.

\begin{figure}[!htb]
\centering
\includegraphics[height=0.245\textwidth]{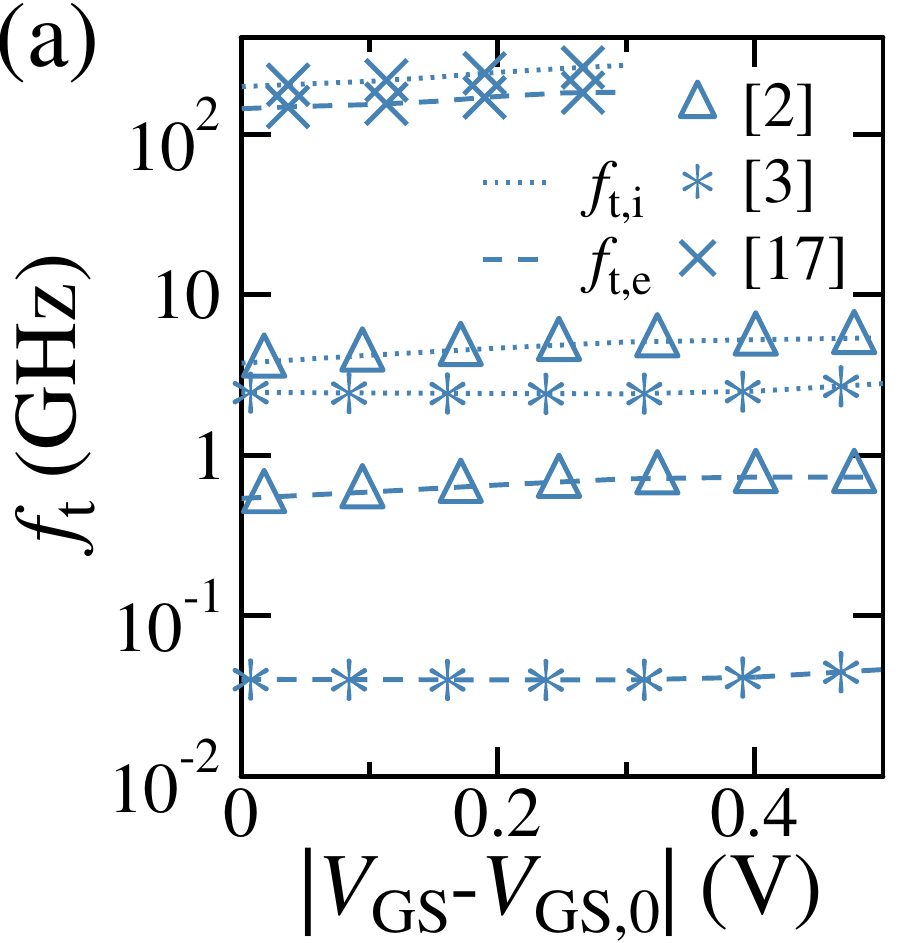}
\includegraphics[height=0.245\textwidth]{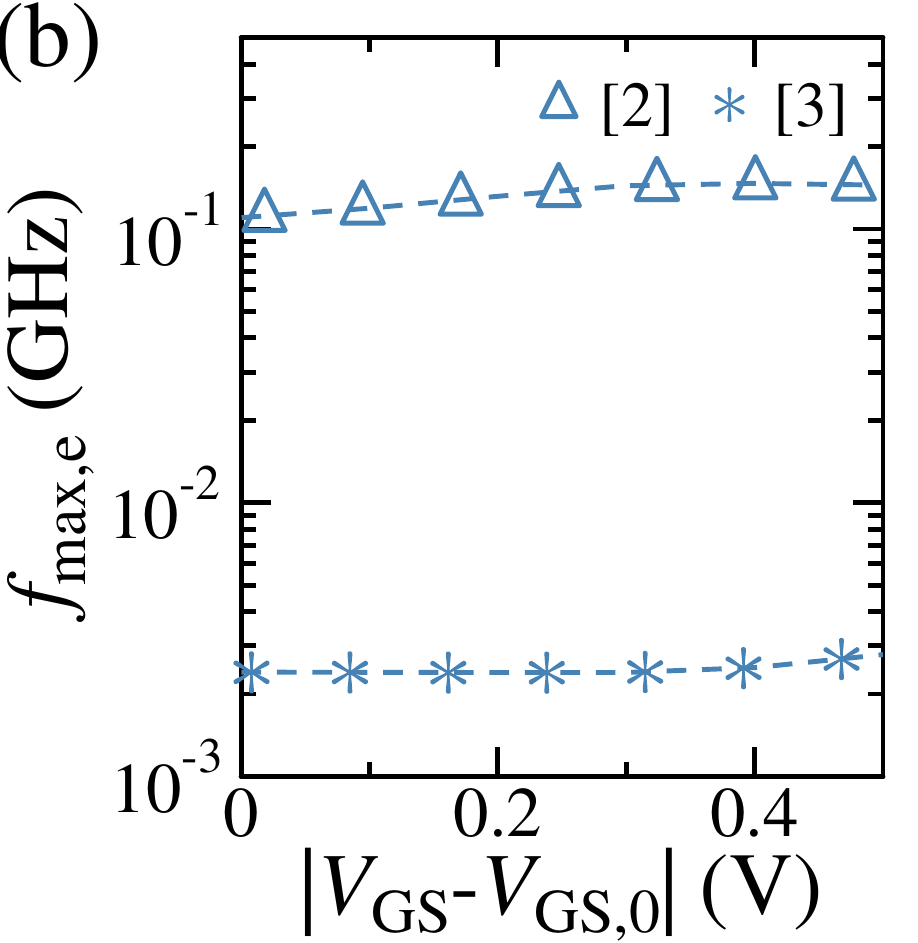} \\
\caption{(a) Intrinsic and extrinsic cutoff frequency and (b) extrinsic maximum oscillation frequency within the bias range where $R_{\rm{C,2}}$ has been extracted for different technologies: \SI{60}{\nano\meter}-long device \cite{WuZou16} ($V_{\rm{DS}}=\SI{0.1}{\volt}$), \SI{100}{\nano\meter}-long device \cite{LyuLu16} ($V_{\rm{DS}}=\SI{0.3}{\volt}$) and \SI{250}{\nano\meter}-long device \cite{GuoDon13} ($V_{\rm{DS}}=\SI{0.5}{\volt}$).}
\label{fig:hf_fom}
\end{figure}

The HF figures of merit defined in Eqs. (\ref{eq:fte}) and (\ref{eq:fmaxe}) have been obtained for a \SI{60}{\nano\meter}-long device \cite{WuZou16}, a \SI{100}{\nano\meter}-long device \cite{LyuLu16} and a \SI{250}{\nano\meter}-long device \cite{GuoDon13} from different technologies, using $R_{\rm{C,2}}$ and the corresponding $C_{\rm{gs,i/gd,i}}$ and $R_{\rm{g}}$ reported in the corresponding reference. $g_{\rm{m,i}}$ and $g_{\rm{d,i}}$ vary also with the bias point according to the reference data. Results are shown in Fig. \ref{fig:hf_fom}. Notice that $f_{\rm{t,e}}$ and $f_{\rm{max,e}}$ of the best HF device reported in each work can not be reproduced here due to the lack of information required for the applied YFM, e.g., transfer characteristic not reported or reported at different bias.

The impact of $R_{\rm{C}}$ can be observed in the difference between intrinsic and extrinsic cutoff frequency: the highest the $R_{\rm{C,2}}$ values (associated to the device in \cite{LyuLu16}), the most degraded $f_{\rm{t,e}}$ with respect to $f_{\rm{t,i}}$. In addition to electrostatic effects and in contrast to mature technologies, $f_{\rm{max,e}}$ in HF GFETs is strongly affected not only by $R_{\rm{C,2}}$ but also by $R_{\rm{g}}$. The latter one can be optimized by T-type gate architectures \cite{GuoDon13}, \cite{WuZou16} while the former requires further optimization which can be evaluated with the extraction methods presented here.

%applied to the \SI{250}{\nano\meter}-long device \cite{GuoDon13}; \SI{100}{\nano\meter}-long device \cite{LyuLu16}; \SI{300}{\nano\meter}-long device \cite{LyuLu16}, \SI{60}{\nano\meter}-long device \cite{WuZou16}
%
%and \cite{LyuLu16}

\section{Conclusion}

Efficient and immediate contact resistance extraction methods, developed within the context of drift-diffusion theory, have been described and applied to graphene FETs from different technologies. In contrast to other technology-specific and more intricate approaches, e.g., AMs, TLM, the extraction methodologies presented here are based on individual and practical transistor static characteristics, i.e., no additional test structures nor a description of internal physical phenomena are required. $V_{\rm{GS}}$-independent and $V_{\rm{GS}}$-dependent contact resistance values, $R_{\rm{C,1}}$ and $R_{\rm{C,2}}$, respectively, can be extracted according to the applied methodology. A drift-diffusion drain current model including the extracted parameters describes the transfer characteristics of the studied devices from different technologies. Extracted values are in good agreement with reference values of different simulation frameworks. Furthermore, the extracted $R_{\rm{C}}$s of fabricated GFET technologies are close to the reference values obtained by other less straightforward methods. Immediate evaluation of contacts is enabled by the methods. $R_{\rm{C,2}}$ enables the evaluation of the contact transparency as well as high-frequency performance projections considering the bias-dependent potential barrier at the metal-channel interface. The methods are expected to be applied to any 2D transistor technology within the bias range in which the carrier transport can be described by a drift-diffusion approach.

%% The Appendices part is started with the command \appendix;
%% appendix sections are then done as normal sections
%% \appendix

%% \section{}
%% \label{}

%% If you have bibdatabase file and want bibtex to generate the
%% bibitems, please use
%%
%%  \bibliographystyle{elsarticle-harv} 
%%  \bibliography{<your bibdatabase>}

%% else use the following coding to input the bibitems directly in the
%% TeX file.

\section*{Acknowledgements}

This project has received funding from the European Union’s Horizon 2020 research and innovation programme under grant agreements No GrapheneCore2 785219 and No GrapheneCore3 881603, from Ministerio de Ciencia, Innovación y Universidades under grant agreement RTI 2018-097876-BC21(MCIU/AEI/FEDER, UE), and project 001-P-0011702-GraphCAT: Comunitat Emergent de grafè a Catalunya, co-funded by FEDER within the framework of Programa Operatiu FEDER de Catalunya 2014-2020.

%% The Appendices part is started with the command \appendix;
%% appendix sections are then done as normal sections
%% \appendix

%% \section{}
%% \label{}

%% If you have bibdatabase file and want bibtex to generate the
%% bibitems, please use
%%
%%  \bibliographystyle{elsarticle-num} 
%%  \bibliography{<your bibdatabase>}

%% else use the following coding to input the bibitems directly in the
%% TeX file.

\end{document}